\documentclass[nofootinbib,preprint,superscriptaddress]{revtex4-1}
\usepackage{amsmath,graphicx,hyperref}

\newcommand{\fms}[1]{{#1}\!\!\!/}

\newcommand{\mc}{\mathcal}
\newcommand{\mr}{\mathrm}

\newcommand{\mO}{\mathcal{O}}

\newcommand{\be}{\begin{equation}} 
\newcommand{\ee}{\end{equation}} 
\newcommand{\bea}{\begin{eqnarray}} 
\newcommand{\eea}{\end{eqnarray}}

\newcommand{\pp}{\perp}
\newcommand{\dg}{\dagger}
\newcommand{\n}{\overline{n}}
\newcommand{\nn}{\frac{\fms{\overline{n}}}{2}}

\newcommand{\bl}[1]{{\bf{#1}}}

\newcommand{\blp}[1]{{\bf{#1}}_{\perp}}
\newcommand{\blpu}[1]{{\bf{#1}}^{\perp}}

\newcommand{\nnb}{\nonumber} 

\newcommand{\as}{\alpha_s} 

\newcommand{\eps}{\epsilon} 
\newcommand{\veps}{\varepsilon} 

\newcommand{\UV}{\eps_{\mr{UV}}}
\newcommand{\IR}{\eps_{\mr{IR}}}

\begin{document}

%\vspace*{18pt}

%%%%%%%%%%%%%%%%%%%%%%%%%%%%%%%%%%%%%%%%%%%%%%%%%%%%%%%%%%%%%%%%%%%%%%
%%%%%%%%%%%%%%%%%%%%%%%%%%%%% Title %%%%%%%%%%%%%%%%%%%%%%%%%%%%%%%%%%
%%%%%%%%%%%%%%%%%%%%%%%%%%%%%%%%%%%%%%%%%%%%%%%%%%%%%%%%%%%%%%%%%%%%%%

\title{Fragmentation to a jet in the large $z$ limit }

\def\Seoultech{Institute of Convergence Fundamental Studies and School of Liberal Arts, Seoul National University of Science and Technology, Seoul 01811, Korea}
\def\Pitt{Pittsburgh Particle Physics Astrophysics and Cosmology Center (PITT PACC) \\ Department of Physics and Astronomy, University of Pittsburgh, Pittsburgh, Pennsylvania 15260, USA}

\author{Lin Dai}
\email[E-mail:]{lid33@pitt.edu}
\affiliation{\Pitt}
\author{Chul Kim}
\email[E-mail:]{chul@seoultech.ac.kr}
\affiliation{\Seoultech\vspace{0.7cm}} 
\author{Adam K. Leibovich}
\email[E-mail:]{akl2@pitt.edu}
\affiliation{\Pitt}

\begin{abstract} \vspace{0.1cm}\baselineskip 3.0ex 

We consider the fragmentation of a parton into a jet with small radius $R$ in the large $z$ limit, where $z$ is the ratio of the jet energy to the mother parton energy.  In this region of phase space, large logarithms of both $R$ and $1-z$ can appear, requiring resummation in order to have a well defined perturbative expansion.  Using soft-collinear effective theory, we study the fragmentation function to a jet (FFJ) in this endpoint region.  We derive a factorization theorem for this object, separating collinear and collinear-soft modes.  This allows for the resummation using renormalization group evolution of the logarithms $\ln R$ and $\ln(1-z)$ simultaneously.  We show results valid to next-to-leading logarithmic order for the global Sudakov logarithms.  We also discuss the possibility of non-global logarithms that should appear at two-loops and give an estimate of their size.  

\end{abstract}

\maketitle 

%%%%%%%%%%%%%%%%%%%%%%%%%%%%%%%%%%%%%%%%%%%%%%%%%%%%%%%%%%%%%%%%%%%%%%
\baselineskip 3.3ex

\section{Introduction} 

The fragmentation function (FF)~\cite{Collins:1981uw}, which describes an energetic splitting of a parton into a final state, is a very important ingredient in understanding high-energy % jet and 
hadron production. Using the FF we can systematically separate short- and long-distance interactions related to the production. For instance, inclusive hadron production for $e^+e^-$ annihilation can be factorized as 
\be\label{eehX} 
\frac{d\sigma(e^+e^-\to hX)}{dE_h} = \int^1_{z_h} \frac{dz}{z} \frac{d\sigma_i (z_h/z,\mu)}{dE_i} D_{h/i}(z,\mu),
\ee
where $i$ denotes the flavor of the produced parton, $z_h = 2E_h/E_{cm}$, and $z=E_h/E_i$. Here, $E_{cm}$ is the center of the mass energy of the collision. The partonic scattering cross section $\sigma_i$ includes the hard interactions for $e^+e^- \to iX$. 
Long-distance interactions describing the fragmenting process from  parton $i$ to  hadron $h$ are encoded in the FF, $D_{h/i}(z)$. 
%Since the FF is given independently of the hard process and can be applied to other scattering processes, it is universal in that sense. 
The FF is universal in the sense that it is independent of the hard process and can be applied to other scattering processes. 
Hence, the FF has long been studied in order to understand its properties. (For details we refer to a recent review~\cite{Metz:2016swz} and the references therein.)

Because we can directly observe a jet using well-defined jet algorithms such as the ones introduced in Refs.~\cite{Catani:1993hr,Ellis:1993tq,Dokshitzer:1997in,Salam:2007xv,Cacciari:2008gp}, it is possible to describe the fragmentation function to a jet (FFJ), as long as the the jet radius, $R$, is enough small~\cite{Dasgupta:2014yra}. 
(For a recent review of jet physics see, for example, \cite{Sapeta:2015gee}.)
Moreover, once the FFJ for the isolated jet is given, we can systematically investigate its substructures (e.g. hadron and subjet fragmentations~\cite{Procura:2011aq,Baumgart:2014upa,Kaufmann:2015hma,Chien:2015ctp,Dai:2016hzf,Kang:2016ehg}, and jet mass~\cite{Idilbi:2016hoa} and transverse momentum~\cite{Bain:2016rrv,Neill:2016vbi} distributions), constructing factorization theorems in connection with the frgmenting jet functions~\cite{Procura:2009vm,Jain:2011xz,Ritzmann:2014mka}.

Analytical results of the FFJ have been calculated up to the next-to-leading order (NLO) in $\as$~\cite{Kaufmann:2015hma,Kang:2016mcy,Dai:2016hzf}. Unlike the hadron FF, the FFJ does not have any infrared (IR) divergence due to  the finite size of the jet radius $R$. However, the presence of large logarithms of $R$ does not give a reliable result in perturbation theory and requires  resummation to all order in $\as$. As shown in Refs.~\cite{Dasgupta:2014yra,Kaufmann:2015hma,Kang:2016mcy,Dai:2016hzf}, resumming logarithms of $R$ is equivalent to running down to a scale $\mu\sim QR$ using Dokshitzer-Gribov-Lipatov-Altarelli-Parisi (DGLAP) evolution equations, where $Q$ is a hard energy comparable to the jet energy, $E_J$. This resummed result of the FFJ has been successfully applied to inclusive jet~\cite{Kang:2016mcy,Dasgupta:2016bnd} and hadron~\cite{Kang:2016ehg} production, where the effects of various values of $R$ have been investigated in detail. 

If we observe a highly energetic jet, we would expect that most of the energetic splitting processes are captured within the jet radius $R$ since these processes favor small angle radiation. This implies that the large $z$ region gives the dominant contribution to the FFJ, where $z$ is the ratio of the jet energy fraction over the mother parton energy. 
Accordingly, in the perturbative result for the FFJ there are large logarithms of $1-z$, which need to be resummed to all order in $\as$. Already at one loop order there appears a double logarithm $\ln(1-z)/(1-z)_+ \sim L^2$, where $L$ schematic represents a large logarithm. At  leading logarithm (LL) accuracy, the  resummed can be represented as $\sum_{k=0} C_k (\as L^2)^{k} \sim \exp (L f_0 (\as L))$, which gives the dominant correction to the perturbative expansion of the FFJ.  
 
Thus, for a proper description of the FFJ in the large $z$ limit, we have to systematically handle large logarithms of $1-z$ as well as large logarithms of $R$. In general, if some quantity involves several distinct scales, we try to factorize it so that each factorized part can be well described at one  properly chosen scale. Then performing evolutions between these largely separated scales, we resum the large logarithms. For the FFJ, soft-collinear effective theory (SCET)~\cite{Bauer:2000ew,Bauer:2000yr,Bauer:2001yt,Bauer:2002nz} provides the appropriate framework for factorization and enable us to resum large logarithms automatically by solving the renormalization group (RG) equations for the factorized parts. 

Near the endpoint where $z\to 1$, the FFJ  consists of dynamics with two well-separated scales. Since an observed jet carries most of energy of the mother parton, radiation outside the jet should be soft  with energy~$\sim E_J(1-z)$. Therefore the jet splitting process can be initiated by soft dynamics, while radiation inside the jet is described dominantly by collinear interactions. However, in the effective theory approach wide angle soft interactions are not adequate for explaining the radiation outside the narrow jet because they cannot effectively recognize the jet boundary characterized by the small radius $R$. Instead, we introduce a more refined soft mode, namely the collinear-soft mode~\cite{Bauer:2011uc,Procura:2014cba}, which can resolve the narrow jet boundary and can consistently describe the lower energy, out-of-jet radiations. The collinear-soft mode has previously been used to factorize the cross sections for a narrow jet at a low energy scale~\cite{Becher:2015hka,Chien:2015cka,Becher:2016mmh,Kolodrubetz:2016dzb}. 

In this paper, using SCET we construct a factorization theorem for the FFJ near the endpoint considering collinear and collinear-soft interactions.\footnote{\baselineskip 3.0ex
In a strict sense our factorization theorem would hold up to NLO in $\as$. Beyond NLO, large nonglobal logarithms~(NGLs)~\cite{Dasgupta:2001sh,Banfi:2002hw} that are sensitive to a restricted jet phase space might appear and require some modification of our factorization theorem presented here.}   
Then we resum the large logarithms of $1-z$ and $R$ simultaneously.
In sec.~\ref{sec2} we discuss the characteristics of large-$z$ physics for the FFJ and factorize it into the collinear and the collinear-soft pieces. Then, we confirm our factorized result through NLO by an explicit calculation of each factorized part. In sec.~\ref{RGs}, based on the factorization, we resum the large logarithms by performing RG evolution. We also discuss  
large nonglobal logarithms (NGLs) that possibly contribute to NLL accuracy. 
In sec.~\ref{nume} the numerical results of the FFJ to the accuracy of NLL plus NLO in $\as$ are shown.
Finally in sec.~\ref{conc} we conclude. 

\section{The FFJ in the limit $z\to 1$}
\label{sec2}

Using SCET, the FFJ can be defined as~\cite{Dai:2016hzf}
\bea
\label{qJFF}
D_{J_k/q}(z,\mu) &=& \sum_{X_{\notin J},X_{J-1}} \frac{z^{D-3}}{2N_c} \mr{Tr} \langle 0 | \delta \Bigl(\frac{p_J^+}{z}-\mc{P}_+\Bigr) \nn \Psi_n | J_k(p_J^+,R) X_{\notin J}\rangle 
\langle J_k(p_J^+,R) X_{\notin J} | \bar{\Psi}_n |0\rangle,\\
\label{gJFF}
D_{J_k/g} (z,\mu) &=&  \sum_{X_{\notin J},X_{J-1}} \frac{z^{D-3}}{p_J^+(D-2)(N_c^2-1)}   \\
&&\times \mr{Tr} \langle 0 |  \delta\Bigl(\frac{p_J^+}{z}-\mc{P}_+\Bigr) \mc{B}_n^{\pp\mu,a}
| J_k(p_J^+,R) X_{\notin J}\rangle 
\langle J_k(p_J^+,R) X_{\notin J} | \mc{B}_{n\mu}^{\pp a}| 0 \rangle_. \nnb
\eea 
Here $\Psi_n=W_n^{\dagger} \xi_n$ and $\mc{B}_{n}^{\pp\mu,a}  = i\n^{\rho}g_{\perp}^{\mu\nu} G_{n,\rho\nu}^b \mc{W}_n^{ba} = i\n^{\rho}g_{\perp}^{\mu\nu} \mc{W}_n^{\dagger,ba} G_{n,\rho\nu}^b$ are gauge invariant collinear quark and gluon field strength respectively. $W_n$ ($\mc{W}_n$) is a collinear Wilson line in the fundamental (adjoint) representation~\cite{Bauer:2000yr,Bauer:2001yt}. These collinear fields have momentum scaling $p_n^{\mu} = (p_+,p_{\pp},p_-) = Q(1,\lambda,\lambda^2)$, where $\lambda$ is a small parameter comparable to small jet radius $R$. $p_{\pm}$ are denoted as $p_+\equiv \n\cdot p = p_0 + \hat{\bl{n}}_J\cdot \bl{p}$ and $p_-\equiv n\cdot p = p_0 -\hat{\bl{n}}_J\cdot \bl{p}$, where $\hat{\bl{n}}_J$ is a unit vector in the jet direction and two lightcone vectors $n^{\mu} = (1,\hat{\bl{n}}_J)$ and $\n^{\mu} = (1,-\hat{\bl{n}}_J)$ have been employed. The expressions for the FFJs in Eqs.~(\ref{qJFF}) and (\ref{gJFF}) are valid in the jet frame where the transverse momentum of the observed jet, $\blpu{p}_J$, is zero.  

In this paper, we will consider inclusive $\mr{k_T}$-type algorithms~\cite{Catani:1993hr,Ellis:1993tq,Dokshitzer:1997in,Cacciari:2008gp}, where the merging condition of two light particles is given by 
\be\label{mergingcon}
\theta < R'.
\ee
Here $\theta$ is the angle between the two particles, and $R'=R$ for an $e^+e^-$ collider and $R'=R/\cosh y$ for a hadron collider, where $y\sim\mO(1)$ is the rapidity for the central region. 

The definitions of the FFJs in Eqs.~(\ref{qJFF}) and (\ref{gJFF}) hold for $z \sim \mO(1)$, but are not reliable near the endpoint where $z$ goes to 1. In the limit $z \to 1$, the observed jet takes most of the energy from the mother parton and hence the jet splitting (out-jet) contributions should be described by soft gluon radiation. If $1-z$ is power counted as $\mO(\eta)$ with $\eta \ll 1$, the relevant soft mode would have momentum scaling $k \sim  (k_+,k_{\pp},k_-) \sim Q(\eta,\eta,\eta)$. However, for the proper resummation of $\ln R$, we need a mode that can probe the jet boundary expressed in terms of $R$. 
This mode would have a lower resolution than the soft mode while the $k_+$ component  should still be power counted as $\mO(\eta)$. Because the jet merging criterion for the soft gluon radiation is given by~\cite{Ellis:2010rwa} 
\be\label{softconst} 
\tan^2 \frac{R'}{2} > \frac{k_-}{k_+}\ , 
\ee
the proper mode should allow for the hierarchy, $k_- \sim k_+ \lambda^2 \ll k_+$, where $\lambda \sim R$. 
Thus this mode should have scaling $k\sim Q \eta (1,\lambda,\lambda^2)$. From now on we will call this mode the collinear-soft mode.  

We can consistently separate the usual soft mode $\sim Q(\eta,\eta,\eta)$ and the collinear-soft mode as was first done in the di-jet scattering cross section~\cite{Becher:2015hka,Chien:2015cka}. Furthermore, the separation of the collinear-soft mode from the collinear fields has been %systematically 
performed in the formulation of $\mr{SCET_+}$~\cite{Bauer:2011uc}. Because the collinear-soft mode can be considered as a subset of the usual soft mode, we have to subtract the overlapped of the collinear-soft contribution from the soft contribution in loop calculations similar to the usual zero-bin subtractions~\cite{Manohar:2006nz}. 

If we apply this process to the FFJ with $z\to 1$, we see that the soft contributions can be cancelled by the collinear-soft subtractions. 
Since the soft mode with a scaling $(k_+,k_-) \sim Q(\eta,\eta)$ cannot resolve the jet boundary in Eq.~(\ref{softconst}), the real soft gluon radiation does not contribute to the in-jet contribution of the FFJ, while the out-jet contribution from real radiation covers the full phase space of $(k_+,k_-)$. 
%Instead the radiations contributes to the jet splitting covering the full phase space of $(k_+,k_-)$. 
Thus, independent of $R$, the total soft contributions will be expressed as a function of $1-z$, namely $S(1-z)$. For the collinear-soft contribution that needs to be subtracted from the soft contribution, we apply the same boundary conditions used for the soft mode. Hence the real collinear-soft radiation have only the out-jet contributions, which are the same as the soft mode. Therefore the net result of the collinear-soft contributions that are  to be subtracted  are the same as $S(1-z)$, canceling the soft contribution. 

Finally we are left with a collinear-soft mode at the lower energy scale. When we apply this to the FFJ, we have to keep the jet boundary constraint in Eq.~(\ref{softconst}). 
As a result the active collinear-soft contributions can be expressed in terms of $1-z$ and $R$ simultaneously. 
As we will see, the one loop collinear-soft contributions involve  double logarithms of $\ln \mu/((1-z)E_JR')$. 
This fact indicates that the collinear-soft interactions are responsible for large logarithms of $1-z$ and its resummation would give the dominant contribution to the FFJ near the endpoint.
%This fact indicates that all order summation such as $\sum_{k=1} \as^k \ln^{2k} (1-z)R$ is crucial for the estimation of the JFF with $1\to z$ and the collinear-soft interactions are responsible for this large logarithms resummation. 

\subsection{Factorization of the FFJ when $z\to 1$}  

With the above reasoning, we can systematically extend the FFJs to the endpoint region including collinear-soft interactions. 
%Following the manner performed in Ref.~\cite{Bauer:2011uc}, 
We first decouple the soft mode~$\sim Q(\eta,\eta,\eta)$ from the collinear mode~$\sim Q(1,R,R^2)$. Then we introduce the collinear-soft mode $\sim Q\eta(1,R,R^2)$ in the collinear sector, classifying collinear and collinear-soft gluons as $A_n^{\mu} \to A_n^{\mu} + A_{n,cs}^{\mu}$. Accordingly the covariant derivative in the collinear sector decomposes as $iD^{\mu} = iD_c^{\mu} +iD_{cs}^{\mu} = \mc{P}^{\mu} + gA_n^{\mu} + i\partial^{\mu} + gA_{n,cs}^{\mu}$, where $\mc{P}^{\mu}~(i\partial^{\mu})$ returns collinear (collinear-soft) momentum. In this decomposition, the commutation relations, $[\mc{P}^{\mu},A_{n,cs}^{\nu}]=[\mc{\partial}^{\mu},A_{n}^{\nu}]=0$, hold.
For the factorization of the FFJ, our strategy is simple: after the decomposition into the collinear and  collinear-soft modes, we first integrate out collinear interactions with $p_c^2 \sim Q^2R^2$. As we shall see, this gives an integrated jet function inside a jet. Then at the lower scale $\mu_{cs} \sim Q\eta R$ we will consider the collinear-soft interactions for the jet splitting.  

As performed in Ref.~\cite{Bauer:2011uc}, at  low energy we can additionally introduce so called `ultra-collinear' modes after integrating out the collinear interactions with offshellness $p_c^2 \sim Q^2R^2$. These modes have  energy of the same order as the collinear mode, but their fluctuations are much smaller than $Q^2R^2$. Then at the low energy scale an external collinear field $\phi(=\xi,A)_n$ would be matched onto the ultra-collinear fields, $\phi_n= \phi_{n_1}+\phi_{n_2}+\cdots$, where the lightcone vectors $n_{i=1,2,\cdots}$ reside inside the jet with radius $R$. Note that collinear interactions between different ultra-collinear modes are forbidden since we have already integrated out the large collinear fluctuations $\sim Q^2R^2$. Moreover, as these ultra-collinear modes  reside within the collinear interactions, they cannot resolve the jet boundary. Therefore their interactions do not contribute to the FFJs, at least to NLO in $\as$. So for simplicity we will not consider ultra-collinear interactions in the FFJ. However, in a  more refined jet observable identifying subjets,   these modes may have to be  included.

Adding the collinear-soft mode, the quark initiated FFJ  can be more generically expressed as 
\be\label{qJFFext} 
D_{J_k/q}(z,\mu) = \sum_{X_{\notin J},X_{J-1}} \frac{z^{D-3}}{2N_c} \mr{Tr} \langle 0 | \delta \Bigl(\frac{p_J^+}{z}-\n\cdot iD\Bigr) \nn \xi_n | J_k(p_J^+,R) X_{\notin J}\rangle \langle J_k(p_J^+,R) X_{\notin J} | \bar{\xi}_n|0\rangle.  
\ee
Compared to Eq.~(\ref{qJFF}),  $W_n \delta(p_J^+/z - \mc{P}_+) W_n^{\dagger} =  \delta(p_J^+/z - \n\cdot iD_c)$ has been replaced with $\delta(p_J^+/z - \n\cdot iD)$ in Eq.~(\ref{qJFFext}). 

In order to satisfy gauge invariances at each order in $\lambda\sim \mO(R)$ and $\eta$, following the procedure considered in Ref.~\cite{Bauer:2003mga}, we redefine the collinear gluon field,
\be\label{reAn} 
A_n^{\mu} = \hat{A}_n^{\mu} + \hat{W}_n [iD_{cs}^{\mu},\hat{W}_n^{\dagger}]_,
\ee
where $\hat{A}_n$ are newly defined collinear gluon fields and $\hat{W}_n$ is the collinear Wilson line expressed in terms of $\hat{A}_n$.
As a consequence the covariant derivative in Eq.~(\ref{qJFFext}) can be rewritten as 
\be\label{recod} 
iD^{\mu} = iD_c^{\mu} + W_n iD_{cs}^{\mu} W_n^{\dagger}, 
\ee 
where collinear fields on the right-hand side are the redefined fields and we removed the hat for simplicity.  
Employing Eq.~(\ref{recod}), the delta function in Eq.~(\ref{qJFFext}) can be rewritten as 
\be
\delta \Bigl(\frac{p_J^+}{z}-\n\cdot iD\Bigr) = W_n \delta \Bigl(\frac{p_J^+}{z}-\mc{P}_+-\n\cdot iD_{cs} \Bigr) W^{\dagger}_{n~.}
\ee

Similar to the decoupling of leading ultrasoft interactions from collinear fields~\cite{Bauer:2001yt}, we can 
remove collinear-soft interactions through the term $g n\cdot A_{cs}$ in the Lagrangian of the collinear sector. 
To accomplish this, the collinear quark and gluon fields can be additionally redefined as 
\be\label{recol}
\xi_n \to Y_n^{cs} \xi_n,~~~A_n^{\mu} \to Y_n^{cs} A_n^{\mu} Y_{n}^{cs\dagger},
\ee
where $Y_n^{cs}$ is the collinear-soft Wilson line that satisfies $n\cdot iD_{cs} Y_n^{cs} = Y_n^{cs} n\cdot i\partial$ and has the usual form~\cite{Bauer:2001yt,Chay:2004zn} 
% if we replace $A_{n,cs}$ with $A_{us}$.
\be\label{Ysc}
Y_n^{cs}(x) = \mr{P}~\exp \Biggl[ig\int^{\infty}_x ds n\cdot A_{cs} (sn)\Biggr]\ . 
\ee

Using Eqs.~(\ref{recod}) and (\ref{recol}) we rewrite Eq.~(\ref{qJFFext}) as 
\bea
D_{J_k/q}(z,\mu) &=& \sum_{X_{\notin J},X_{J-1}} \frac{z^{D-3}}{2N_c} \mr{Tr} \langle 0 | \delta \Bigl(\frac{p_J^+}{z}-\mc{P}_+-i\partial_+\Bigr) \nn Y_{\n}^{cs\dagger}Y_n^{cs} W_n^{\dagger} \xi_n | J_k(p_J^+,R) X_{\notin J}\rangle \nnb \\
\label{qJFFext1} 
&&\qquad \times  \langle J_k(p_J^+,R) X_{\notin J} | \bar{\xi}_n W_n Y_n^{cs\dagger}Y_{\n}^{cs}|0\rangle_,  
\eea
where we used the relation $\n\cdot iD_{cs} = Y_{\n}^{cs} i\partial_+ Y_{\n}^{cs\dg}$ and $Y_{\n}^{cs}$ has the same form as Eq.~(\ref{Ysc}) with replacement of $n\to \n$. We also used the crossing symmetry $\phi \cdots |X_{\phi}\rangle = \langle X_{\phi} | \cdots \phi$, where $\phi = W_n,~Y_{\n}^{cs}$. The FFJ in Eq.~(\ref{qJFFext1}) can describe regions of ordinary $z\sim {\cal O}(1)$ and $z\to 1$. If $z$ is ordinary and not too close to 1, we can suppress $i\partial_+$ in the argument of the delta function, since $p_J^+/z -\mc{P}_+ \sim \mO(Q)$ is power counted much larger than $i\partial_+\sim \mO(Q\eta)$. Thus the collinear-soft Wilson lines  cancel by unitarity and we recover the form in Eq.~(\ref{qJFF}). However, when $z\to1$, $p_J^+/z - \mc{P}_+$ becomes the same size as $i\partial_+$, and  we cannot ignore the term $i\partial_+$ in the delta function, which gives nonzero contributions of collinear-soft interactions.

Since $\mc{P}_+$ returns collinear (label) momentum in Eq.~(\ref{qJFFext1}), $\mc{P}_+$ can be fixed as $p_J^+$ near the endpoint. Further, it means that collinear interactions are relevant only for jet merging (in-jet) contribution to the FFJ. Therefore the FFJ in the limit $z\to 1$ can be expressed as\footnote{\baselineskip 3.0ex
Note that the splitting $q\to J_g$ in the limit $z\to 1$ is power suppressed by $\mO(1-z)$ compared to the splitting $q\to J_q$. For $q\to J_g$, the splitted parton away from the observed jet is the collinear-soft quark, which gives a power suppression of $\mO(\eta)$ compared to the collinear-soft gluon radiation. Similarly, for gluon splitting, $g \to J_g$  dominants for the same reason. 
} 
\bea
D_{J_q/q}(z\to 1,\mu) &=& \sum_{X_{\notin J},X_{J-1}} \frac{z^{D-3}}{2N_c} \mr{Tr} \langle 0 | Y_{\n}^{cs\dagger}Y_n^{cs} \nn W_n^{\dagger} \xi_n | J_q(p_J^+,R) X_{\notin J}\rangle \nnb \\
&& \qquad\times  \langle J_q(p_J^+,R) X_{\notin J} | \bar{\xi}_n W_n \delta \Bigl(\frac{p_J^+}{z}-\mc{P}_+^{\dg}+i\partial_+\Bigr)Y_n^{cs\dagger}Y_{\n}^{cs}|0\rangle  \nnb \\
&=&  \sum_{X_c\in J} \frac{1}{2N_c} \mr{Tr} \langle 0 | \nn W_n^{\dagger} \xi_n | q X_c \in J\rangle 
\langle q X_c \in J | \bar{\xi}_n W_n |0 \rangle 
\cdot \sum_{X_{cs}} \frac{1}{N_c} \mr{Tr} \langle 0 | Y_{\n}^{cs\dagger}Y_n^{cs} |X_{cs} \rangle
\nnb \\
\label{qJFFE} 
&&\qquad\times %\sum_{X_{sc}} \frac{1}{N_c} \mr{Tr} \langle 0 | Y_{\n}^{sc\dagger}Y_n^{sc} |X_{sc} \rangle
\langle X_{cs} | \delta\bigl((1-z)p_J^+ + \Theta(\theta -R')i\partial_+ \bigr) Y_n^{cs\dagger}Y_{\n}^{cs}|0\rangle_, 
%\nnb \\
%&=& \mc{J}_q(E_J R',\theta<R') S_q (z;(1-z) E_J R').
\eea
where $\Theta$ is the step function and we reorganized the final states into collinear states $(qX_c)$ in the jet and collinear-soft states $X_{cs}$ in order to factorize collinear and collinear-soft interactions.
In the second equality we fixed the collinear label momentum $\mc{P}^{\dagger}$ as $p_J^+$, and then we put the jet splitting constraint in front of $i\partial_+$ because only the out-jet collinear-soft radiation gives a nonzero contribution for the region $z < 1$. From Eq.~(\ref{softconst}), the jet splitting constraint~$\Theta(\theta -R')$ is equivalent to $\tan^2 R'/2 < k_-/k_+$, where $k$ is the collinear-soft momentum. 

Eq.~(\ref{qJFFE}) shows that the quark FFJ in the limit $z\to 1$ is factorized as 
\be\label{JFFefact}
D_{J_q/q}(z\to 1,\mu;E_JR',(1-z)E_JR') 
= \mc{J}_q(\mu;E_J R',\theta<R') \cdot S_q (z,\mu;(1-z) E_J R'), 
\ee
where $\mc{J}_q$ is the integrated jet function for the in-jet contribution, defined as 
\be\label{injetf} 
\mc{J}_q(\mu;E_J R',\theta<R') =\sum_{X_c \in J} \frac{1}{2N_c~p_J^+} \mr{Tr} \langle 0 | \nn W_n^{\dagger} \xi_n | q X_c \in J(E_J,R')\rangle 
\langle q X_c \in J | \bar{\xi}_n W_n |0 \rangle_.
\ee
$S_q$ is the dimensionless collinear-soft function. When we rewrite $S_q = p_J^+ \tilde{S}_q$, the dimensionful collinear-soft function $\tilde{S}_q$ can be expressed as 
\be\label{SCqf}
\tilde{S}_q (\ell_+,\mu;\ell_+ t) = \sum_{X_{cs}} \frac{1}{N_c} \mr{Tr} \langle 0 | Y_{\n}^{cs\dagger}Y_n^{cs} |X_{cs} \rangle
\langle X_{cs} | \delta\bigl(\ell_+ + \Theta(\theta -R')i\partial_+ \bigr) Y_n^{cs\dagger}Y_{\n}^{cs}|0\rangle_,
\ee
where $t\equiv \tan R'/2$, and $\ell_+ t$ is the scale that will minimize large logarithms in the higher order corrections, as we will see later. 

Using the adjoint representation and taking a similar procedure as we did with the quark case, we obtain the factorization formula for the gluon FFJ,
\be\label{JFFefactg}
D_{J_g/g}(z\to 1,\mu) 
= \mc{J}_g(\mu;E_J R',\theta<R') \cdot S_g (z,\mu;(1-z) E_J R'), 
\ee
where $\mc{J}_g$ is the gluon integrated jet function, and $S_g$ is the collinear-soft function defined similar to Eq.~(\ref{SCqf}), with the Wilson lines in the adjoint representation replacing $Y_{n,\n}^{cs}$. 

\subsection{NLO calculation of the FFJ near the endpoint} 

The integrated jet functions shown in Eqs.~(\ref{JFFefact}) and (\ref{JFFefactg}) have been explicitly computed at NLO~\cite{Cheung:2009sg,Ellis:2010rwa,Chay:2015ila} and partially computed at NNLO~\cite{Chien:2015cka,Becher:2016mmh}. The NLO results with the constraint of Eq.~(\ref{mergingcon}) read 
\bea\label{qintj} 
\mc{J}_q(\mu;E_J R',\theta<R') &=& 1+\frac{\as C_F}{2\pi} \Biggl[\frac{1}{\UV^2}+ \frac1{\UV}\Bigl(\frac32 + \ln\frac{\mu^2}{p_J^{+2} t^2}\Bigr)\nnb\\
&&~~~~~~~~~~~
+ \frac{3}{2}\ln\frac{\mu^2}{p_{J}^{+2}t^2}+\frac{1}{2}\ln^2\frac{\mu^2}{p_{J}^{+2}t^2}+\frac{13}{2}-\frac{3\pi^2}{4} \Biggr]\ , \\
\mc{J}_g(\mu;E_J R',\theta<R') &=& 1+\frac{\as C_A}{2\pi} \Biggl[\frac{1}{\UV^2}+\frac{1}{\UV}\Bigl(\frac{\beta_0}{2C_A} +\ln\frac{\mu^2}{p_{J}^{+2}t^2}\Bigr) +\frac{\beta_0}{2C_A}\ln\frac{\mu^2}{p_{J}^{+2}t^2}\nnb \\ 
\label{gintj}
&&~~~~~~~~~~~
+\frac{1}{2}\ln^2\frac{\mu^2}{p_{J}^{+2}t^2}+\frac{67}{9}-\frac{23n_f}{18C_A}-\frac{3\pi^2}{4} \Biggr]\ ,
\eea
where $p_J^+ t \sim E_J R'$, $\beta_0 = 11N_c/3-2n_f/3$, $C_A=N_c=3$, and $n_f$ is the number of flavors. 

For the NLO computation of the collinear-soft function in Eq.~(\ref{SCqf}) we consider virtual and real gluon contributions respectively. Separating ultraviolet (UV) and infrared (IR) divergences carefully, the virtual contributions are given by 
\be\label{vsoft} 
M_V^S = -\frac{\as C_F}{\pi} \Bigl(\frac{1}{\UV}-\frac{1}{\IR}\Bigr)^2 \delta(\ell_+).
\ee
The real contributions at one loop can be written as 
\bea
M_R^S &=& \frac{\as C_F}{\pi} \frac{(\mu^2 e^{\gamma_E})^{\eps}}{\Gamma(1-\eps)} \int^{\infty}_0 dk_+ dk_- (k_+k_-)^{-1-\eps}
\Bigl[\delta(\ell_+-k_+) \Theta(k_--t^2k_+) \nnb \\
&&+\delta(\ell_+) \Theta(t^2k_+ - k_-)\Bigr]\equiv M_{R1}^S +  M_{R2}^S, 
\eea
where $k$ is the momentum of the outgoing collinear-soft gluon and $M_{R1}^S~(M_{R2}^S)$ indicates the contribution from the first (second) term in the square brackets. 

\begin{figure}[t]
\begin{center}
\includegraphics[height=9cm]{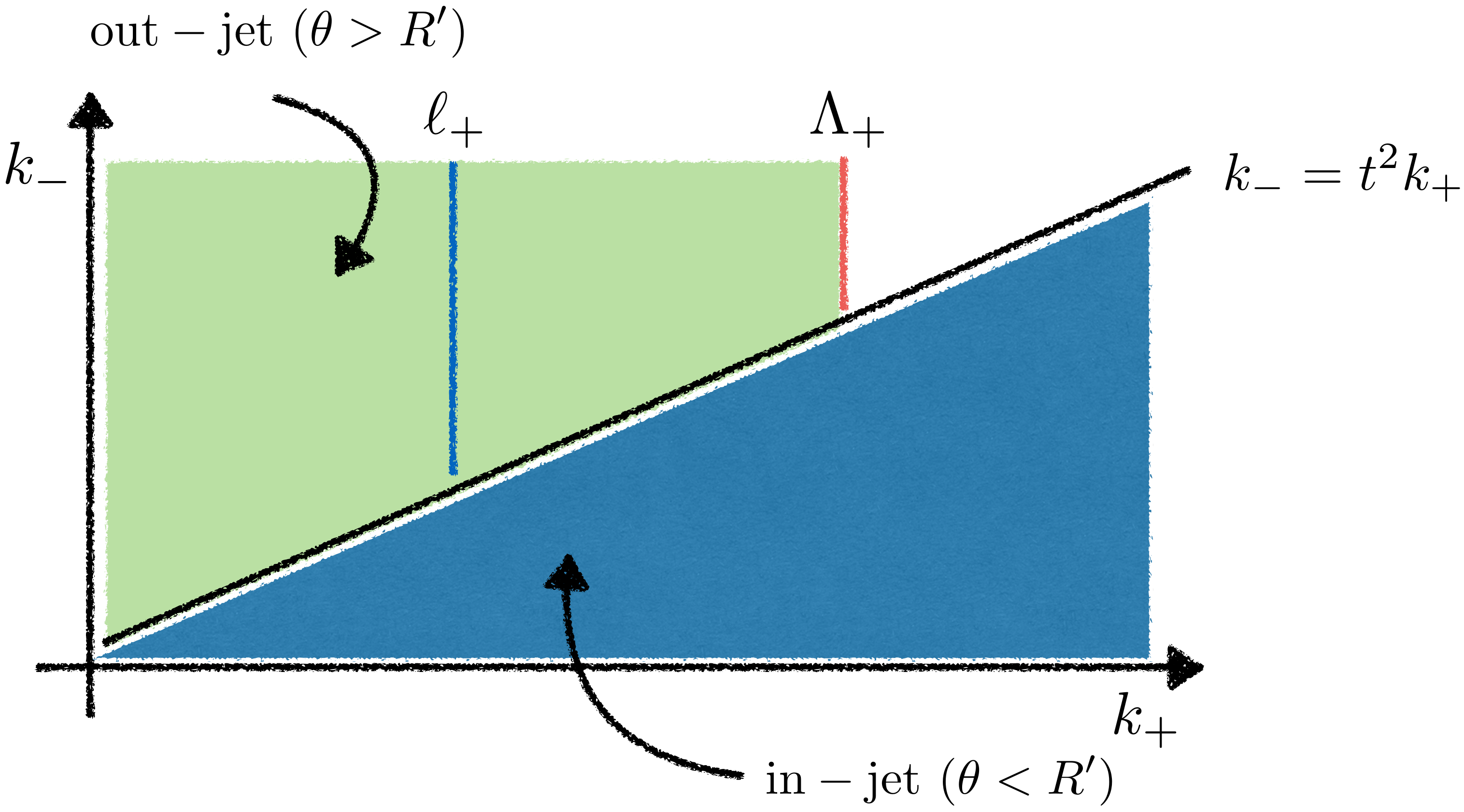}
\end{center}
\vspace{-0.3cm}
\caption{\label{scregion} \baselineskip 3.0ex Phase space for the real gluon emission in the collinear-soft function. In the $(k_+,k_-)$ plane, the region above the border line $k_-=t^2k_+$ gives the out-jet contribution and the region below gives the in-jet contribution. 
$\Lambda_+$ is the maximum value for the distribution of $\ell_+$ and can be chosen arbitrarily.}
\end{figure}

In Fig.~\ref{scregion} we show the possible phase space for the emitted collinear-soft gluon after the integration over $\blp{k}$.
$M_{R2}^S$ covers  region below the jet border line $(k_-=t^2 k_+)$. Hence the result is
\bea 
M_{R2}^S &=&
\frac{\as C_F}{\pi} \frac{(\mu^2 e^{\gamma_E})^{\eps}}{\Gamma(1-\eps)}\delta(\ell_+) \int^{\infty}_0 dk_+\int^{t^2 k_+}_0 dk_-(k_+k_-)^{-1-\eps} \nnb \\
\label{softR2} 
&=& \frac{\as C_F}{2\pi}\Biggl[\left(\frac{1}{\veps_{\mr{UV}}}-
\frac{1}{\veps_{\mr{IR}}}\right)^2 - \left(\frac{1}{\veps_{\mr{UV}}}-
\frac{1}{\veps_{\mr{IR}}}\right) \ln t^2\Biggr]\delta(\ell_+)_. 
\eea

For $M_{R1}^S$, $k_+$ is fixed to be $\ell_+$ by the delta function, and the possible phase space has been denoted as a blue line in the upper plane in Fig~\ref{scregion}. However we need to extract the IR divergences as $\ell_+ \to 0$. In order to do so, we introduce the so called $\Lambda_+$-distribution, which is defined as  
\be\label{L+dist}
\int_0^{L} d\ell_+[g(\ell_+)]_{\Lambda_+} f(\ell_+)  = \int_0^{L}d\ell_+ g(\ell_+) f(\ell_+) -\int_0^{\Lambda_+} d\ell_+ g(\ell_+) f(0),
\end{equation}
where $f(\ell_+)$ is an arbitrary smooth function at $\ell_+ = 0$. $\Lambda_+$ is an arbitrary upper limit for $\Lambda_+$-distribution and is power counted to have the same size as $\ell_+$. 
We can write $M_{R1}^S$ using this distribution, 
\bea
M_{R1}^{S} &=& \frac{\as C_F}{\pi} \frac{(\mu^2 e^{\gamma_E})^{\eps}}{\Gamma(1-\eps)}
~\ell_+^{-1-\eps} \int^{\infty}_{t^2\ell_+} dk_- k_-^{-1-\eps} \nnb \\
\label{MR1}
&=& \delta(\ell_+) I_{R1}(\Lambda_+, t)  + \frac{\as C_F}{\pi} \frac{(\mu^2 e^{\gamma_E})^{\eps}}{\Gamma(1-\eps)}
\Biggl[\ell_+^{-1-\eps} \int^{\infty}_{t^2\ell_+} k_-^{-1-\eps} \Biggr]_{\Lambda_+}\ , 
\eea 
where the integration region for $I_{R1}$  corresponds to the green region in Fig.~\ref{scregion}. Integrating over this region, we get
\bea
I_{R1} &=& \frac{\as C_F}{\pi} \frac{(\mu^2 e^{\gamma_E})^{\eps}}{\Gamma(1-\eps)} 
\Biggl[\int^{\infty}_0 dk_+\int^{\infty}_{t^2k_+} dk_-
(k_+k_-)^{-1-\eps} -\int^{\infty}_{\Lambda_+} dk_+\int^{\infty}_{t^2 k_+} dk_-(k_+k_-)^{-1-\eps}\Biggr] \nnb \\
\label{IR1}
&=& \frac{\as C_F}{2\pi} \Biggl[
 \left(\frac{1}{\eps_{\mr{UV}}}-
\frac{1}{\eps_{\mr{IR}}}\right)^2 + \left(\frac{1}{\eps_{\mr{UV}}}-
\frac{1}{\eps_{\mr{IR}}}\right) \ln t^2 \\
&&~~~~~~-\left(\frac{1}{\UV^2}+\frac{1}{\UV}\ln\frac{\mu^2}{\Lambda_+^2 t^2}+\frac{1}{2}\ln^2\frac{\mu^2}{\Lambda_+^2t^2}-\frac{\pi^2}{12}\right)\Biggr]\ . \nnb
\eea 
The second term in Eq.~(\ref{MR1}) is given by
\be
\label{MR1dist}
\frac{\as C_F}{\pi} \frac{(\mu^2 e^{\gamma_E})^{\eps}}{\Gamma(1-\eps)}
\Biggl[\ell_+^{-1-\eps} \int^{\infty}_{t^2\ell_+} k_-^{-1-\eps} \Biggr]_{\Lambda_+}
= \frac{\as C_F}{\pi} \Biggl[\frac{1}{\ell_+} \Bigl(\frac{1}{\UV}+\ln\frac{\mu^2}{\ell_+^2t^2}\Bigr)\Biggr]_{\Lambda_+}\ . 
\ee

Finally combining Eqs.~(\ref{vsoft}), (\ref{softR2}), (\ref{IR1}) and (\ref{MR1dist}) we obtain the bare one loop result of $\tilde{S}_q$,
\bea
M_S &=& M_V^S + M_{R1}^S + M_{R2}^S \nnb \\
\label{bareS}
&=& \frac{\as C_F}{\pi} \Biggl\{\delta(\ell_+)\Biggl(-\frac{1}{2\UV^2}-\frac{1}{2\UV}\ln\frac{\mu^2}{\Lambda_+^2 t^2}-\frac{1}{4}\ln^2\frac{\mu^2}{\Lambda_+^2t^2}+\frac{\pi^2}{24}\Biggr) 
\\
&&~~~~~~~~~~+\Biggl[\frac{1}{\ell_+} \Bigl(\frac{1}{\UV}+\ln\frac{\mu^2}{\ell_+^2t^2}\Bigr)\Biggr]_{\Lambda_+} \Biggr\}\ .\nnb
\eea
The one loop result of the collinear-soft function for gluon FFJ is the same if we replace $C_F$ with $C_A=N_c$ in Eq.~(\ref{bareS}).

Since the dimensionless soft-collinear function, $S_{k=q,g}(z)=p_J^+ \tilde{S}_{k} (\ell_+)$, is a function of $z$, we need to express the $\Lambda_+$-distribution in terms of the standard plus distribution of $z$. From Eq.~(\ref{L+dist}) we obtain the relation
\be
\label{reldist}
[\tilde{g}(\ell_+)]_{\Lambda_+} = \frac{1}{p_J^+} [g(z)]_+ +\frac{1}{p_J^+} \delta(1-z) \int^{b}_0 dz' g(z'),  
\ee
where $\ell_+ = p_J^+(1-z)$ and $g(z) = p_J^+\tilde{g}(\ell_+)$. In the $\Lambda_+$-distribution, $\Lambda_+$ has been replaced with $p_J^+(1-b)$, where $b$ is a dimensionless parameter close to 1.

Finally, the dimensionless collinear-soft functions at NLO can be written as follows: 
\bea
S_{k=q,g}(z,\mu;(1-z)E_JR') &=& \delta(1-z) + \frac{\as C_k}{2\pi} \Biggl\{\delta(1-z)\Bigl(-\frac{1}{2} \ln^2 \frac{\mu^2}{p_J^{+2}t^2}+\frac{\pi^2}{12}\Bigr) \nnb \\
\label{SqgNLO} 
&&~~~~~~~~~~~~+2\Bigl[\frac{1}{(1-z)}\Bigl(\ln \frac{\mu^2}{p_J^{+2}t^2} - 2\ln(1-z) \Bigr)\Bigr]_+\Biggr\}\ ,
\eea
where $C_q = C_F$ and $C_g=C_A$. As can be seen in Eq.~(\ref{SqgNLO}), the scale necessary to minimize the large logarithms in the collinear-soft functions is  $(1-z)E_JR'$. In the limit $z\to 1$, running the collinear-soft function will be required to obtain a precise estimate of the FFJ. 

In Eqs.~(\ref{JFFefact}) and (\ref{JFFefactg}) we have shown the factorization theorem near the endpoint. Combining Eqs.~(\ref{qintj}), (\ref{gintj}) and (\ref{SqgNLO}) we can easily check that the fixed NLO results of Eqs.~(\ref{JFFefact}) and (\ref{JFFefactg}) recover the NLO results of FFJs for the full range~\cite{Dai:2016hzf,Kaufmann:2015hma,Kang:2016mcy} when we take the limit $z\to 1$. 

\section{Renormalization Group Evolution and Resummation of Large Logarithms}
\label{RGs}

\subsection{RG evolution from the factorization of the FFJ}
\label{RGf}
Based on the factorized results in Eqs.~(\ref{JFFefact}) and (\ref{JFFefactg}), we can systematically resum the large logarithms of  $\ln R$ and $\ln(1-z)$ in the FFJ using the RG evolutions of the integrated jet function $\mc{J}_k$ and the collinear-soft functions $S_k$. The FFJ in the limit $z\to 1$ can be factorized at an arbitrary factorization scale $\mu_f$. Then $\mc{J}_k$ can be evolved from $\mu_f$ to collinear scale $\mu_c \sim E_JR'$, where the large logarithms at the higher order in $\as$ are minimized and the perturbative expansion is safely convergent. Simultaneously we can evolve $S_k$ from $\mu_f$ to $\mu_{cs} \sim (1-z) E_J R'$ to minimize the large logarithms at $\mu_{cs}$. 
%Because the fixed order results of the integrated jet function at $\mu_c$ and the collinear-soft function at $\mu_{cs}$ do not involve large logarithms, RG evolutions from $\mu_f$ to $\mu_c$ and $\mu_{cs}$ can automatically include resum whole large logarithms, and the final result can be expressed as an exponentiation form of the large logarithms.    

The anomalous dimensions of the integrated jet functions and the collinear-soft functions defined by 
\bea
\label{defanoj} 
\frac{d}{d\ln\mu} \mc{J}_{k} (\mu) &=& \gamma_{c,k} (\mu) \mc{J}_{k}(\mu), \\
\label{defanocs} 
\frac{d}{d\ln\mu} S_{k} (x,\mu) &=& \int^1_x\frac{dz}{z} \gamma_{cs,k}(z,\mu) S_{k}(x/z,\mu), 
\eea
where $k=q,~g$, are obtained from Eqs.~(\ref{qintj}), (\ref{gintj}), and (\ref{SqgNLO}) at one loop,
\bea
\label{anojLO}
\gamma_{c,q}^{(0)} &=& \frac{\as C_F}{2\pi} \Bigl(2\ln \frac{\mu^2}{E_J^2 R^{'2}} +3\Bigr)\ ,~~~
\gamma_{c,g}^{(0)} = \frac{\as C_A}{2\pi} \Bigl(2\ln \frac{\mu^2}{E_J^2 R^{'2}} +\frac{\beta_0}{C_A} \Bigr) , \\
\label{anocsLO}
\gamma_{cs,k}^{(0)} (z) &=& \frac{\as C_k}{2\pi} \Bigl(-2\ln \frac{\mu^2}{E_J^2 R^{'2}} \delta(1-z) + \frac{4}{(1-z)_+}\Bigr)\ ,
\eea 
where $p_J^+ t$ is approximated as $E_JR'$. When we combine Eqs.~(\ref{anojLO}) multiplied by $\delta(1-z)$ and Eq.~(\ref{anocsLO}), the logarithmic terms  cancel and the well-known DGLAP splitting kernels in the limit $z\to 1$ are reproduced:
\be
\delta(1-z) \gamma_{c,k}^{(0)} + \gamma_{cs,k}^{(0)} (z) = \frac{\as}{\pi} P_{kk}^{(0)} (z\to 1) .
\ee

Logarithmic terms in the leading anomalous dimensions indicate the presence of the cusp anomalous dimension. Beyond LL accuracy, the anomalous dimensions can be expressed as 
\bea
\label{anoj}
\gamma_{c,k} &=& A_c \Gamma_{C,k} (\as) \ln \frac{\mu^2}{E_J^2R^{'2}} + \hat{\gamma}_{c,k} (\as), \\
\label{anocs}
\gamma_{cs,k} (z) &=& \delta(1-z) \Bigl[A_{cs} \Gamma_{C,k} (\as) \ln \frac{\mu^2}{E_J^2R^{'2}}+\hat{\gamma}_{cs,k} (\as) \Bigr]
-\kappa_{cs} A_{cs} \frac{\Gamma_{C,k} (\as)}{(1-z)_{+}}\ ,
\eea
where $\Gamma_{C,k} = \sum_{n=0} \Gamma_{n,k}(\as/4\pi)^{n+1} $ are the cusp anomalous dimensions obtained from calculations of  light-like Wilson loops~\cite{Korchemsky:1987wg,Korchemskaya:1992je}. The first two coefficients are given by 
\be
\Gamma_{0,k} = 4C_k,~~~\Gamma_{1,k} = 4C_k \Biggl[\Bigl(\frac{67}{9}-\frac{\pi^2}{3}\Bigr) C_A - \frac{10}{9} n_f\Biggr]\ .
\ee
From the LO results in Eqs.~(\ref{anojLO}) and (\ref{anocsLO}) we extract $\{A_c,A_{cs},\kappa_{cs}\} = \{1,-1,2\}$ and the noncusp anomalous dimensions $\hat{\gamma}_{c,q}=3\as C_F/(2\pi) + \mO(\as^2)$, $\hat{\gamma}_{c,g}=\as \beta_0/(2\pi) + \mO(\as^2)$, and  $\hat{\gamma}_{cs,k}= \mO(\as^2)$.

Using Eqs.~(\ref{anoj}) and (\ref{anocs}) we perform RG evolutions of the integrated jet functions and the collinear-soft functions up to next-to-leading logarithmic (NLL) accruarcy. For $\mc{J}_k$ the result of the RG evolution from $\mu_f$ to $\mu_c$ can be written as 
\be
\label{RGj}
\mc{J}_k (\mu_f) = \exp \Bigl[2 A_c S_{\Gamma} (\mu_f,\mu_c) + A_c \ln \frac{\mu_f^2}{E_J^2R^{'2}} a[\Gamma_{C,k}](\mu_f,\mu_c)
+a[\hat{\gamma}_{c,k}](\mu_f,\mu_c)\Bigr]\mc{J}_k (\mu_c).
\ee
Here $S_{\Gamma}$ and $a[f]$ are 
\be
S_{\Gamma} (\mu_f,\mu_c) = \int^{\alpha_f}_{\alpha_c} \frac{d\as}{b(\as)} \Gamma_{C,k}(\as) \int^{\as}_{\alpha_f} \frac{d\as'}{b(\as')},~~~a[f](\mu_f,\mu_c) = \int^{\alpha_f}_{\alpha_c} \frac{d\as}{b(\as)} f(\as),
\ee
where $\alpha_{f,c} \equiv \as (\mu_{f,c})$ and $b(\as) = d\as/(d\ln\mu)$ is QCD beta function.

For the evolution of $S_k$, following the conventional method introduced in Refs.~\cite{Neubert:2005nt,Becher:2006nr}, we obtain 
\bea
\label{RGcs}
S_{k}(z,\mu_f) &=& \exp \Bigl[2 A_{cs} S_{\Gamma} (\mu_f,\mu_{cs}) +a[\hat{\gamma}_{cs,k}](\mu_f,\mu_{cs})\Bigr] \left(\frac{\mu_f^2}{E_J^2R^{'2}}\right)^{-\eta_S/\kappa_{cs}} \\
&&~~~\times \bar{S}_k \Bigl[\ln \frac{\mu_{cs}^2}{E_J^2R^{'2}} -2\partial_{\eta_S}\Bigr] \frac{e^{-\gamma_E \eta_S}}{\Gamma(\eta_S)} (1-z)^{(-1+\eta_S)}\ , \nnb
\eea
where $\eta_S$ is defined as $\eta_S = -\kappa_{cs}A_{cs} a[\Gamma_{C,k}](\mu_f,\mu_{cs})$ and is positive  for $\mu_f > \mu_{cs}$. $\bar{S}_k$ is 
\be
\bar{S}_k [L] = 1+ \frac{\as C_k}{2\pi} \Bigl(-\frac{1}{2} L^2 - \frac{\pi^2}{4} \Bigr) + \mO(\as^2).
\ee

\subsection{Contribution of nonglobal logarithms} \label{NGL-sec}

When we extend the factorized result of the FFJ to the two loop or higher order in $\as$, one important issue is the presence of nonglobal logarithms (NGLs)~\cite{Dasgupta:2001sh,Banfi:2002hw}. Usually  NGLs appear when jet observables cover a limited phase space due to the jet algorithm and arises from multiple gluon radiations near the jet boundary. Especially when there are large energy differences  between in-jet and out-jet radiated gluons,  large NGLs are unavoidable. 

For the FFJ near the endpoint there are two modes that could resolve the jet boundary and give nonvanishing contributions.  
The collinear mode with large energy certainly radiates only inside a jet, but the collinear-soft mode can radiate across a jet boundary and give a nonvanishing result as  $z\to 1$ at the lower energy scale. So we conjecture there can exist  large NGLs in the FFJ in the large $z$ limit.

In order to systematically resum  large NGLs, we would need to modify our  factorization theorem as it is  designed to resum global Sudakov logarithms. To include resummation of  NGLs using effective theory, at two loop order we might have to consider dressed collinear-soft gluons decoupled from a (ultra-)collinear gluon along a certain direction inside a jet, which could give rise to a new dipole operator other than $Y_{n,\n}^{cs}$ at  low energy. We will not pursue such a refined factorization theorem here, but we mention that some advanced treatments of  NGLs have been recently introduced in Refs.~\cite{Becher:2015hka,Becher:2016mmh,Caron-Huot:2015bja,Larkoski:2015zka,Neill:2015nya,Larkoski:2016zzc,Becher:2016omr}.

To estimate the size of the NGLs in the FFJ, we note that they should have same form as the endpoint logarithms, $\ln(1-z)$, which can be inferred from the  ratio of scales between the  collinear scale~$\mu_c \sim E_JR'$ and the collinear-soft scale $\mu_{cs}\sim (1-z)ER'$. As seen in the threshold expansion of inclusive jet production~\cite{deFlorian:2013qia}, leading NGLs start to appear at two loops, $\as^2 L^2 \sim \as^2 (\ln(1-z)/(1-z))_+$, where $L$ schematically denotes a large logarithm. So at NLL accuracy we have to resum these leading NGLs to all order in $\as$, i.e.,  $\sum_{n=2} C_{\rm{NG}}^n (\as L)^n$.
 
For the hemisphere jet mass distribution in $e^+e^-$ annihilation, the resummed result of leading NGLs is known in the large $N_c$ limit~\cite{Dasgupta:2001sh}. Interestingly the resummed result of leading NGLs for an individual narrow jet is found to have the same form as the case of the hemisphere jet mass,  the only difference simply arising from the need to choose suitable evolution scales~\cite{Banfi:2010pa,Dasgupta:2012hg}. Therefore, using the result in Ref.~\cite{Dasgupta:2001sh} we conjecture the resummed result of leading NGLs for the FFJ in the large $N_c$ limit should be of the form
\be\label{RNGL}
\Delta^k_{\mr{NG}} (\mu_{c},\mu_{cs}) = \exp\Biggl(-C_A C_k \frac{\pi^2}{3} \Bigl(\frac{1+(at)^2}{1+(bt)^c}\Bigr) t^2 \Biggr)\ ,
\ee
where $k=q,~g$, and 
\be
t=\frac{1}{\beta_0} \ln{\frac{\as(\mu_{cs})}{\as(\mu_{c})}} \sim -\frac{1}{\beta_0}\ln \Bigl(1-\frac{\beta_0}{4\pi} \as(\mu_c) \ln \frac{\mu_c^2}{\mu_{cs}^2}\Bigr)\ .
\ee 
The fit parameters from the Monte Carlo implementation of the parton-shower are given by $a=0.85 C_A,~b=0.86C_A$, and $c=1.33$~\cite{Dasgupta:2001sh}.
Note that our treatment of large logarithm to NLL accuracy only holds for the anti-$\mr{k_T}$ algorithm. As discussed in Ref.~\cite{Banfi:2010pa}, for other $\mr{k_T}$-type algorithms, such as $\mr{k_T}$ and C/A, clustering effects~\cite{Banfi:2005gj,Delenda:2006nf} give rise to additional large logarithmic terms, which can be also present at NLL order.  

Up to NLL accuracy (plus NLO) in $\as$, the resummation factor for NGLs in Eq.~(\ref{RNGL}) just multiplies the resummed results of the FFJ from the previous section, where the resummed expressions of $\mc{J}_{k=q,g}$ and $S_{k=q,g}$ are shown in Eqs.~(\ref{RGj}) and (\ref{RGcs}) respectively. 
In the next section we show various numerical results for the FFJ in the large $z$ region comparing the results using only DGLAP evolutions and our resummed results of the large logarithms as well as the NGLs. 

\section{Numerical Results} 
\label{nume}

\begin{figure}[h]
\begin{center}
\includegraphics[width=16cm]{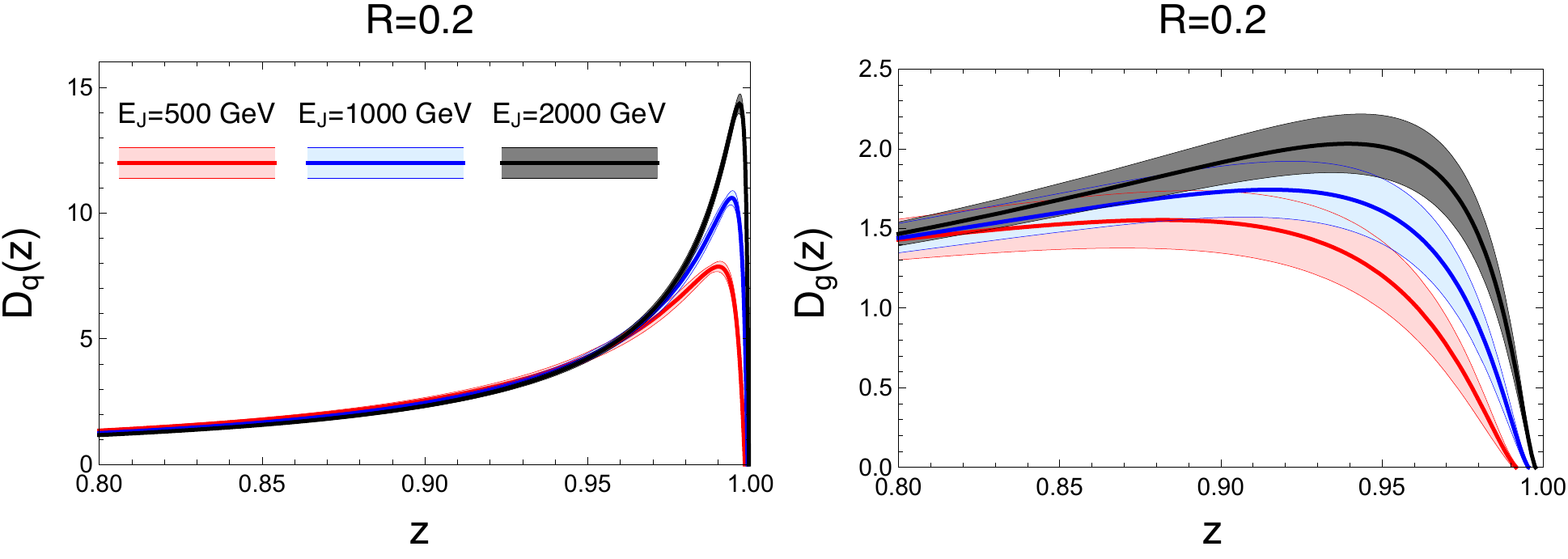}
\end{center}
\vspace{-0.6cm}
\caption{\label{E-fig} \baselineskip 3.0 ex
$D_{J_q/q}(z)$ (left panel) and $D_{J_g/g}(z)$ (right panel) with different jet energies. Red, blue, and black curves correspond to jet energy $E_J$  equal to 500, 1000, and 2000 GeV, respectively. The jet radius is chosen to be $R=0.2$ and the factorization scale is $\mu_f=E_J$. Error estimation is described in the text.}
\end{figure}

\begin{figure}[h]
\begin{center}
\includegraphics[width=16cm]{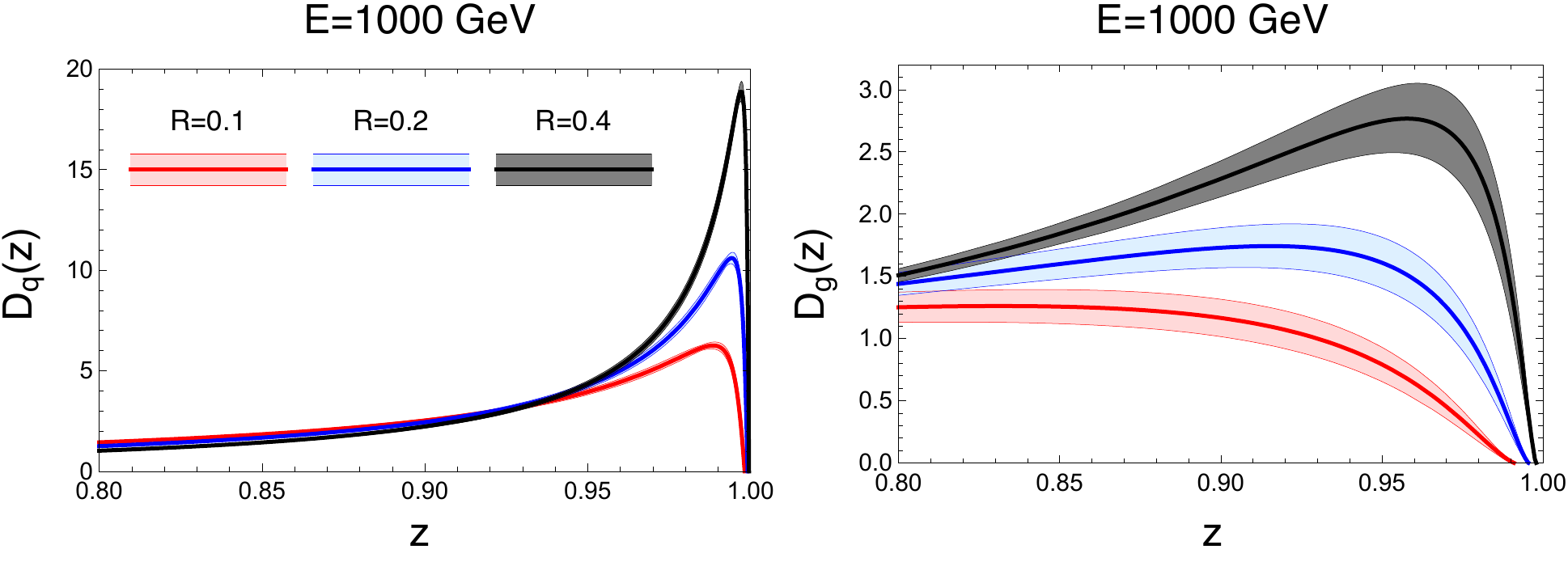}
\end{center}
\vspace{-0.6cm}
\caption{\label{R-fig} \baselineskip 3.0 ex
$D_{J_q/q}(z)$ (left panel) and $D_{J_g/g}(z)$ (right panel) with different jet radii. Red, blue, and black curves correspond to the jet radius $R$ equal to $0.1$, $0.2$, and $0.4$ respectively. The jet energy is $E_J=1000$ GeV and the factorization scale is $\mu_f=E_J$. }
\end{figure}

In this section we show numerical results of the resummed FFJ focusing on the large $z$ region. For simplicity we set $R'=R$. 
As shown in sec.~\ref{RGf}, in order to resum large logarithms in $D_{J_{k}/k}(z,\mu_f)$, the integrated jet functions ${\cal J}_{k}$ are run from the jet scale $\mu_c=ER$ to $\mu_f$, and the collinear-soft functions $S_{k}$ from  $\mu_{cs}=ER(1-z)$ to $\mu_f$. 
Because the FFJ is dependent upon the scale $\mu_f$ (actually following DGLAP evolution), the shape of the FFJ varies for different choices of $\mu_f$. For convenience we choose $\mu_f = E_J$ throughout this section.  Error estimations of the jet and the collinear-soft functions are obtained by varying the jet scale and the collinear-soft scale within $(\mu_c/2, 2\mu_c)$ and $(\mu_{cs}/2, 2\mu_{cs})$ respectively. Then errors of $D_{J_{k}/k}(z,\mu_f)$ are obtained by summing these in quadrature.

Based on the factorized expressions in Eqs.~(\ref{JFFefact}) and (\ref{JFFefactg}), Fig.~\ref{E-fig} shows $\mr{NLL_G+NLO}$ results of $D_{J_q/q}$ and $D_{J_g/g}$ for different energies of jets with the same radius $R=0.2$. Here $\mr{NLL_G}$ represents the NLL accuracy including only global logarithms from the factorization approach in sec.~\ref{RGf}. 
%($E_J$ equal to $500$, $1000$ and $2000$ GeV, respectively), for all of which jet radius is fixed to be $R=0.2$. 
For the extreme endpoint region where $\mu_{cs}=E_JR(1-z) \approx \Lambda_{\rm QCD}$, our description is not reliable because of nonperturbative contributions.  
%Although our focus in this paper is investigating the behavior of FFJs in the large $z$ limit, we should not trust the resummed results when $z$ gets too close to $1$ where non-perturbative effect dominates, i.e. $\mu_{cs}=E_JR(1-z)$ gets close to $\Lambda_{\rm QCD}$. Note that $D_{J_q/q}(z,\mu)$ and $D_{J_g/g}(z,\mu)$ are not observables (both scale dependent), in Fig.~\ref{E-fig} (as well as in Fig.~\ref{R-fig}) we have chosen the common scale $\mu=E_J$. Eventually, one needs to incorporate FFJs in expressions of cross sections which are supposed to be independent of the common scale $\mu$ and could involve other scales such as hard scales and ultra-soft scales as well, depending on the specific processes considered. 
Fig.~\ref{R-fig} shows $\mr{NLL_G+NLO}$ results of $D_{J_q/q}$ and $D_{J_g/g}$ for different jet radii 
%($R$ equal to $0.1$, $0.2$, $0.4$), for all of which jet energy is fixed to be $E_J=1000$ GeV. 
with the jet energy fixed to be 1000~GeV.
From Figs.~\ref{E-fig} and \ref{R-fig} we can see the tendencies that energetic parton showering processes are captured more in the jet as the jet energy $E_J$ and/or the radius $R$ become larger.  

%Note that the characteristic scales of $\mu_c$ and $\mu_{cs}$ both depend on $E_J R$, while Fig.~\ref{E-fig} and Fig.~\ref{R-fig} seem to show that the dependences on $E_J$ and $R$ are different. This is because we have chosen the common scale as $\mu=E_J$. So in Fig.~\ref{E-fig}, all scales $\mu$, $\mu_c$ and $\mu_{cs}$ are changed when we change $E_J$, while in Fig.~\ref{R-fig}, $\mu_c$ and $\mu_{cs}$ are changed with $\mu$ fixed when we change jet radius $R$. 

\begin{figure}[t]
\begin{center}
\includegraphics[width=16cm]{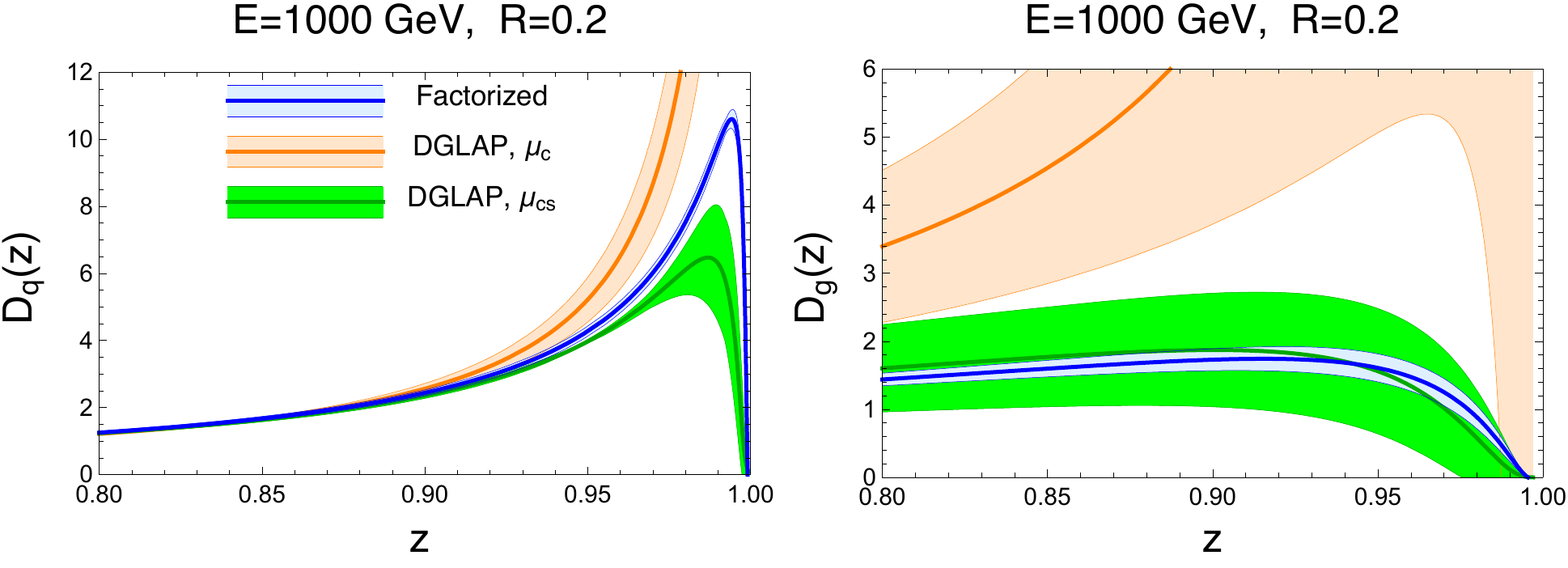}
\end{center}
\vspace{-0.6cm}
\caption{\label{DGLAP-fig} \baselineskip 3.0ex
Comparison of the result using leading DGLAP evolution and the resummed result at $\mr{NLL_G+NLO}$ from the factorization approach. The orange (green) curves are obtained using leading DGLAP evolution with FFJs running from $\mu_c=E_JR~(\mu_{cs}=E_JR(1-z))$ to $\mu_f = E_J$.  
%The green curves are obtained using DGLAP evolution with FFJs running from $\mu_{cs}=E_JR(1-z)$. 
Blue curves are the resummed result of the FFJs. The jet radius is $R=0.2$ and jet energy is $E_J=1000$ GeV. }
\end{figure}

To see the importance of the factorization description on the FFJs, in Fig.~\ref{DGLAP-fig} we compare the resummed results at $\mr{NLL_G+NLO}$ and the results using leading DGLAP evolution naively.
%from NLO result in the fixed $\as$. 
Here using only DGLAP evolution from $\mu_c = E_JR$ to $\mu_f = E_J$ can be understood as resumming only large logarithms of $R$. 
As $z$ goes to 1, the resummed results of only $\ln R$ blow up. However, when we do DGLAP evolution from $\mu_{cs} = E_JR(1-z)$ to $\mu_f = E_J$, we can see more realistic results. Compared with our factorization approach with the accuracy of $\mr{NLL_G+NLO}$, both DGLAP evolved results involve much larger uncertainties. 
%If we expand Eq.~(\ref{JFFefact}) and keep terms up to order ${\cal O}(\alpha_s)$,  it returns to Eq.~(24) and Eq.~(30)  with $z\to1$ in \cite{Dai:2016hzf}. $D_{J_q/q}(\mu,z)$ and $D_{J_g/g}(\mu,z)$ then satisfy the usual DGLAP equations and can be easily solved analytically using, say, Laplace transformation.

\begin{figure}[h]
\begin{center}
\includegraphics[width=16cm]{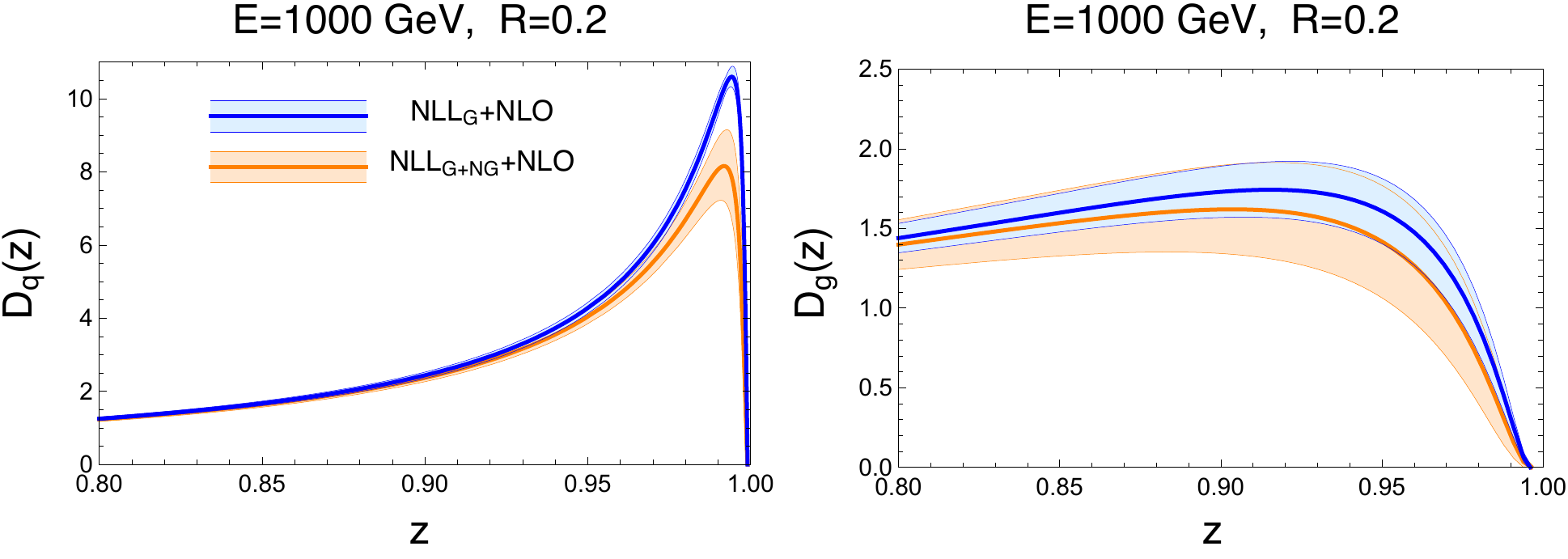}
\end{center}
\vspace{-0.6cm}
\caption{\label{NG-fig} \baselineskip 3.0ex
Comparison of the resummed results with (orange) and without (blue) resumming the NGLs. Here $R=0.2$ and $E_J=1000$ GeV. }
\end{figure}

Fig.~\ref{NG-fig} shows the resummed result of the FFJs with the accuracy of $\mr{NLL_{G+NG}+NLO}$ using our conjectured result for including leading NGLs discussed in sec.~\ref{NGL-sec}, obtained by multiplication of the FFJ at $\mr{NLL_G+NLO}$ by $\Delta^{k=q,g}_{\mr{NG}} (\mu_{c},\mu_{cs})$ in Eq.~(\ref{RNGL}). The result including leading NGLs gives rise to some suppression to the FFJs. 
A similar suppression can be also seen in the light jet mass distribution for the hemisphere jet production when the resummed results including the NGLs are compared with the case without the NGLs~\cite{Becher:2016omr}. 
Because of additional dependences on both $\mu_{c}$ and $\mu_{cs}$ from $\Delta^{k=q,g}_{\mr{NG}} (\mu_{c},\mu_{cs})$, the result  with NGLs increases the errors. The errors might be reduced if we include the NNLO result in $\as$, which is beyond the scope of our paper. 

\begin{figure}[h]
\begin{center}
\includegraphics[height=6cm]{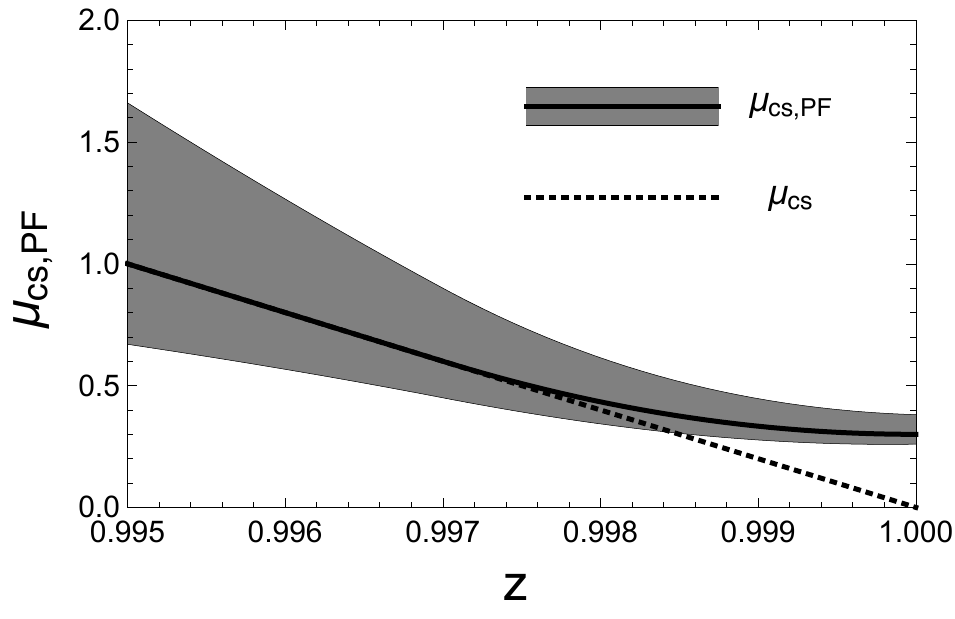}
\end{center}
\vspace{-0.6cm}
\caption{\label{profile-fig} 
\baselineskip 3.0exProfile function $\mu_{cs,PF}$ [solid black curve and gray band, defined in Eq.~(\ref{profile-eq})] is used to estimate errors due to variation of the collinear-soft scale. The dashed line is the $z$ dependent collinear-soft scale $\mu_c(1-z)$ with $\mu_c=200$ GeV.}
\end{figure}

There is one more comment about error estimations used above. Since $\mu_{cs}=E_JR(1-z)$ is $z$ dependent and bound to hit the Landau pole as $z \to 1$, we have used the following profile function to avoid the Landau pole:
% estimate the errors due to collinear-soft scale variation (ref. Fig.~\ref{profile-fig}):
\begin{equation}
\label{profile-eq}
  \mu_{cs,PF}(z) =(1+\frac{\delta}{1+{\rm exp}[(z-z_1)/(1-z_1)]})
    \begin{cases}
      (1-z)\mu_c & \text{if }  z< z_1\\
      \mu_{\rm Min}+a(1-z)^2 & \text{if }  z \geq z_1
    \end{cases} ,
\end{equation}
where $\mu_{\rm Min}=0.3$, $\mu_c=E_J R$, $a$ and $z_1$ are fixed by requiring that $\mu_{cs,PF}(z)$ and its first derivative are continuous at $z=z_1$.  The profile function is shown in Fig.~\ref{profile-fig}.  To vary the collinear-soft scale, we used $\delta=\{0,-0.5,1\}$.  $\mu_{cs,PF}(z)$ is devised to ensure that the collinear-soft scale freezes as it approaches the Landau pole and coincide with $\mu_{cs}(z)$ otherwise.

\section{Conclusion and Outlook}
\label{conc}

In this paper, as shown in Eqs.~(\ref{JFFefact}) and (\ref{JFFefactg}), we have developed a factorization theorem of the FFJ with a small jet radius $R$ in the large $z$ limit. At the scale $\mu\sim E_J R'$ we first integrate out collinear modes with offshellness $p_{c}^2 \sim (E_J R')^2$, and obtain the integrated jet functions, $\mc{J}_{q,g}$. At the lower scale $\mu \sim (1-z)E_JR'$ the collinear-soft mode can probe the jet boundary  and gives a nonvanishing result at  higher order in $\as$. Combining NLO results of the integrated jet function and the collinear-soft function, we can successfully reproduce NLO result of the FFJ in the limit $z\to 1$.

Performing RG evolutions of the factorized jet and collinear-soft functions we resummed large logarithms of $1-z$ and $R$ simultaneously. 
The anomalous dimensions of each factorized function involves the cusp anomalous dimension, which enables us to systematically resum large logarithms
% such as $\ln(1-z)$ 
beyond  leading order. As a result we have shown the resummed result at NLL, which significantly modifies the large $z$ behavior of the FFJ when compared to the result of only resuming logarithms of $R$ through naive DGLAP evolution. Large NGLs may appear at NNLO in $\as$ and could contribute to the resummed result at NLL accuracy. We therefore have  estimated NGL contributions to the FFJ applying the resummed formalism in the large $N_c$ limit~\cite{Dasgupta:2001sh}.

The finite size of the jet radius $R$ plays an important role in performing successful RG evolution of the FFJ in the large $z$ limit.\footnote{\baselineskip 3.0ex
Compared to a massless jet, some differences of a jet with small and finite $R$ have been discussed in the resummation of threshold logarithms~\cite{deFlorian:2013qia,Kidonakis:1998bk}.}
%\footnote{A similar role of small $R$ has been considered in the resummation of threshold logarithms~\cite{deFlorian:2013qia,Kidonakis:1998bk,deFlorian:2007fv}.}  
Even though $R$ is small, the radius  makes it possible to have an observed jet with nonzero invariant mass and each factorized function for the FFJ is IR finite. Similar results occur for the heavy quark fragmentation function~(HQFF) in the large $z$ limit, where the HQFF can be factorized into the heavy quark function and the soft shape function~\cite{Neubert:2007je,Fickinger:2016rfd}. 
Due to a nonzero heavy quark mass $M$,  both functions are IR finite and systematic RG evolutions to the scales $M$ and $M(1-z)$ can be done. 

Note that the FFJ reduces to a light hadron fragmentation function if $R$ goes to zero. In this case the factorization to collinear and collinear-soft interactions breaks down because the relevant anomalous dimensions blow up and RG evolutions become nonperturbative, as can be checked from Eqs.~(\ref{anojLO}) and (\ref{anocsLO}). A similar result can be applied to the parton distribution function (PDF) near the endpoint. Actually, in order to resum large logarithm $\ln(1-z)$ in the PDF, a similar factorization approach to ours has been considered in Ref.~\cite{Fleming:2012kb}, where soft gluon radiation is responsible for the parton splitting. Interestingly the factorized collinear and soft functions for the PDF contains  rapidity divergences~\cite{Chiu:2011qc,Chiu:2012ir} as well as UV and IR divergences. However the rapidity RG evolution turns out to be IR sensitive and become nonperturbative. (We checked if there exist  rapidity divergences in the factorized functions for the FFJ, but the finite size of $R$ forbids  rapidity divergences and guarantees ordinary RG evolutions from pure UV divergences.) 

Our factorized and resummed result analyzed here can be widely applied for energetic jet productions. The resumming procedure of large logarithms of $1-z$ from the effective theory approach can be used for systematic resummations of threshold logarithms for inclusive jet~\cite{deFlorian:2013qia,deFlorian:2007fv} and dijet production~\cite{Kidonakis:1998bk,Hinderer:2014qta}. However, for more precisely resummed results of large logarithms, explicit calculations beyond NLO are required, and a thorough analysis of factorization including NGLs  is needed. 

\acknowledgments

CK was supported by Basic Science Research Program through the National Research Foundation of Korea (NRF) funded by the Ministry of Science, ICT and Future Planning (Grants No. NRF-2014R1A2A1A11052687). AL and LD were supported in part by NSF grant PHY-1519175.

%%%%%%%%%%%%%%%%%%%%%%%%%%%%%%%%%%%%%%%%%%%%%%%%%%%%%%%%%%%%%%%%%%%%%%
%%%%%%%%%%%%%%%%%%%%%%%%%%%%% Bibliography %%%%%%%%%%%%%%%%%%%%%%%%%%%
%%%%%%%%%%%%%%%%%%%%%%%%%%%%%%%%%%%%%%%%%%%%%%%%%%%%%%%%%%%%%%%%%%%%%%

%%%%%%%%%%%%%%%%%%%%%%%%%%%%%%%%%%%%%%%%%%%%%%%%%%%%%%%%%%%%%%%%%%%%%%

\end{document}